\documentclass{article}

\usepackage{amssymb,relsize}
\usepackage{amsthm,amsmath}
\usepackage{color}
\usepackage{graphicx}
\usepackage{caption}
\usepackage{subcaption}
\usepackage[round,sort]{natbib}

\begin{document}


\title{Differential expression analysis for multiple conditions}


\author{Ciaran Evans, Johanna Hardin, Mark Huber, Daniel Stoebel, and Garrett Wong}

\maketitle



\begin{abstract}
As high-throughput sequencing has become common practice, the cost of sequencing large amounts of genetic data has been drastically reduced, leading to much larger data sets for analysis.  One important task is to identify biological conditions that lead to unusually high or low expression of a particular gene.  Packages such as DESeq implement a simple method for testing differential signal when exactly two biological conditions are possible.  For more than two conditions, pairwise testing is typically used. Here the DESeq method is extended so that three or more biological conditions can be assessed simultaneously.
Because the computation time grows exponentially in the number of conditions, a Monte Carlo approach provides a fast way to approximate the $p$-values for the new test. The approach is studied on both simulated data and a data set of {\em C. jejuni}, the bacteria responsible for most food poisoning in the United States.
\end{abstract}





\section{Introduction}

The cost of sequencing large amounts of genetic data continues to
decrease rapidly, leading to new sequencing technologies
able to produce vast quantities of data.  New statistical and computational
techniques are needed to analyze this wealth of new data
(\cite{wanglh2010},\cite{lomanmdcgwp2012}).

In high-throughout sequencing, the genetic material to be analyzed is cut up into millions of small fragments, which are then individually
sequenced.  These sequenced fragments, called {\em reads}, can then be mapped back to the target genome and used for differential expressed
analysis, which compares the relative expression levels of each gene under different biological conditions.  Software packages such as DESeq \cite{andersh2010} and
edgeR \cite{robinsonms2010} are commonly used to determine whether there is a statistically significant difference in expression levels between biological conditions for each gene.

Both DESeq and edgeR are designed to deal with exactly two conditions, but investigators often have three or more such conditions to deal with. For instance, experimenters might be interested in comparing expression of a gene under three concentrations of a protein, at 0\%, 50\%, or 100\%;  in multiple different cell or tissue types in a single organism; from different types of chemical exposures (or different doses of the same chemical); or from a collection of genetic variants.  This leads experimenters with three or more conditions to analyze their results in pairwise fashion.

This work extends the test used by DESeq to three or more biological
conditions to allow simultaneous comparison.  There is a reason DESeq
only uses two conditions:  the computational effort required
to compute the $p$-values for the test grows quickly with the
number of biological conditions.  Therefore a Monte Carlo approach to
estimating the $p$-values will be used.

\subsection{The DESeq Model}

The raw data for relative gene expression is the number of reads for a gene
under each biological condition, usually with multiple samples.
Let $K_{ij}$ denote the number of reads in sample $j$ that are
assigned to gene $i$.  If genes and samples were independent, this
would follow a multinomial distribution, which could then be
approximated by the Poisson distribution.

However, the Poisson model predicts smaller variation than is seen
in the data~\cite{nagalakshmiwwsrgs2008,robinsons2007}, leading
some to assume a negative binomial model, which is used in both the DESeq package~\cite{andersh2010} and
the edgeR package~\cite{robinsonms2010}.

Our notation follows~\cite{andersh2010}.  The model is
\begin{equation}
K_{ij} \sim \textsf{NB}(\mu_{ij},\sigma_{ij}^2),
\end{equation}
where $\mu_{ij}$ and $\sigma_{ij}^2$ are respectively
the mean and variance of the distribution.

Usually $\mu_{ij}$ and $\sigma_{ij}$ are unknown:  in~\cite{andersh2010}
the means are modeled using an expression parameter $q_{i j}$ that represents
the probability a sample from condition $j$ expresses gene $i$.  If
$s_j$ is the number of reads from sample $j$, and $\rho(j)$ maps the
sample $j$ to its biological condition, then
\begin{equation}
\mu_{ij} = q_{i\rho(j)} s_j.
\end{equation}

In order to reduce the number of parameters, the variance is then
modeled as a function of the mean and a known function
$\nu:[0,1] \rightarrow [0,\infty)$
Then the variance becomes
\begin{equation}
\sigma_{ij}^2 = \mu_{ij} + s_j^2 \nu(q_{i\rho(j)}).
\end{equation}

Suppose there are $m$ different biological conditions
$A_1,\ldots,A_m$.  The goal is to test for if the biological condition
has an effect (either positive or negative) on gene expression.
The first statistic needed is the total counts for each condition.
So for $A \in \{A_1,\ldots,A_m\}$, let
\begin{equation}
K_{iA} = \sum_{j:\rho(j) = A} K_{ij}.
\end{equation}

Given $m$ different biological conditions,
the null hypothesis is $H_0:q_{iA_1} = \cdots = q_{i A_m} = q_{i0}$, that is,
the strength of expression does not depend on the biological condition.

Suppose the data (the number of reads for gene $i$ given condition $A$)
is denoted $k_{iA}$.
Start with the pooled mean estimate for $q_{i0}$:
\begin{equation}
\label{EQN:pooledmean}
\hat q_{i0} = \sum_{j} k_{i j} / s_j.
\end{equation}
This in turn yields estimates for each $\mu_{i A}$
\begin{equation}
\label{EQN:meanandvarianceestimator}
\hat \mu_{i A} = \sum_{j:\rho(j) \in A} s_j \hat q_{i0}, \quad
\hat \sigma^2_{iA} = \sum_{j:\rho(j) \in A} s_j \hat q_{i0} +
  s_j^2 \nu(\hat q_{i0}).
\end{equation}

Then the test statistic of Robinson and
Smyth~\cite{robinsons2007} for two conditions is
\begin{equation}
T_i(k_{iA_1},k_{iA_2}) =
  \mathbb{P}(K_{iA_1} = k_{iA_1},K_{iA_2} = k_{iA_2} |
    K_{iA_1} + K_{iA_2} = k_{iA_1} + k_{iA_2}).
\end{equation}

Let $p_i(a_1,a_2) = \mathbb{P}(K_{iA_1} = a_1,K_{iA_2} = a_2)$.  Then the
above test statistic gives a $p$-value that is
\begin{equation}
p_i = \frac{\mathlarger{\sum}\limits_{\substack{a_1 + a_2 = k_{iA_1} + k_{iA_2} \\
    p_i(a_1,a_2) \leq p_i(k_{iA_1},k_{iA_2})}} p_i(a_1,a_2)}
      {\mathlarger{\sum}\limits_{a_1 + a_2 = k_{iA_1} + k_{iA_2}} p_i(a_1,a_2)}
\end{equation}

Since the summation allows for $k_{iA_1} + k_{iA_2} + 1$ different values
of $a_1$, computing each sum exactly takes time $\Theta(k_{iA_1} + k_{iA_2})$.

\section{Extending DESeq for more than two conditions}

Note that the pooled mean estimate \eqref{EQN:pooledmean} can easily
be extended from two conditions to $m > 2$.  With $\hat q_{i0}$ in hand,
the mean and variance estimators \eqref{EQN:meanandvarianceestimator}
remain the same.  Then the test statistic naturally extends to
more than three conditions.
\begin{align*}
T_i(k_{iA_1},\ldots,k_{iA_m}) &=
  \mathbb{P} K_{iA_1} = k_{iA_1},\ldots,K_{iA_m} = k_{iA_m} | S)\\
S &=  \{K_{iA_1} + \cdots + K_{iA_m} = k_{iA_1} + \cdots + k_{iA_m}\}.
\end{align*}

The natural extension of the $p$-value then becomes:
\begin{equation}
p_i = \frac{\mathlarger{\sum}\limits_{\substack{a_1 + \cdots + a_m = k_{iA_1} + \cdots +
    k_{iA_m} \\
    p(a_1,\ldots,a_m) \leq p(k_{iA_1},\ldots,k_{iA_m})}} p(a_1,\ldots,a_m)}
      {\mathlarger{\sum}\limits_{a_1 + \cdots + a_m = k_{iA_1} + \cdots +
   k_{iA_m}} p(a_1,\ldots,a_m)}
\end{equation}

However, the time for computation of the exact $p$-value has grown
tremendously.
Consider $m = 3$.  Instead of a linear number of $(a_1,a_2)$ that
sum to $k_{iA_1} + k_{iA_2}$, there is a quadratic number of nonnegative
integers $(a_1,a_2,a_3)$ that sum to $k_{iA_1} + k_{iA_2} + k_{iA_3}$.

More generally, let $k_{iS} = k_{iA_1} + \cdots + k_{iA_m}$.  Then
the number of nonnegative integers $(a_1,\ldots,a_m)$
that sum to $S$ is well known to be
\begin{equation}
\binom{k_{iS} + m - 1}{m - 1} = \Theta(k_{iS}^{m - 1}).
\end{equation}
Typical data sets can have $k_{iS}$ on the order of $10^6$, making exact computation infeasible
even for small $m$.  Note that the dataset analyzed below has a median total count value of 53,530; over 400 genes with total counts larger than  161,700; and a maximum total count of 23,550,000 (see Table \ref{totcount.tab}).

It is in this case that Monte Carlo methods can be used to estimate the
exact $p$-value (see~\cite{besagc1989,guot1992}).  The idea is simple:
draw vectors $(a_1,\ldots,a_m)$ uniformly conditioned on
$a_1 + \cdots + a_m = k_{iS}$.  This can be accomplished with $m - 1$ random
choices as follows.  Draw a subset of size $m - 1$ from the numbers
$1,\ldots,k_{iS} + m - 1$.  Order the subset, and call it
$b_1 < b_2 < \cdots < b_{m - 1}$.  Let $b_0 = 0$, $b_{m} = k_{iS} + m - 1$, and set
$a_i = b_i - b_{i-1} - 1$ for all $i$ from 1 to $m$.
Then it is easy to verify that the distribution
of the $(a_1,\ldots,a_m)$ is uniform over nonnegative integer vectors
that sum to $k_{iS} + m - 1$.

Now the procedure for estimating the exact $p$-value for gene $i$
is as follows.  Draw $N$ different nonnegative vectors
$(a_1,\ldots,a_m)$ that sum
to $k_{iS}$.  Calculate $p_i(a_1,\ldots,a_m)$ for each such vector,
and let $\hat p_i$ be the proportion of these values that is at
most $p_i(k_{iA_1},\ldots,k_{iA_m})$.  This gives an unbiased estimate
of the exact $p$-value, and standard methods can be used to obtain
confidence intervals.

\section{Simulated Data}

We first simulate data to demonstrate the improved power and false discovery rate of our method.  Using the RNASeq negative binomial simulation model from voom \cite{voom}, we simulate data for 1000 genes from nine samples (three at each of three different conditions).  For 100 genes, we let the first two conditions have the same population distribution.  For the third condition, we set the expression value to be 2-fold different from the first two conditions (i.e., 100 truly DE genes).  For two fold differences, the simulation sets probabilities associated with generating reads from a negative binomial model to be twice as high for one group as compared to another.  The remaining 900 genes are set so that all three conditions come from the same population (i.e., 900 null genes).

\subsection{Power Analysis}

For each of 1000 genes, we ran both the 3-way Monte Carlo simulation method as well as 3 pairwise comparisons of DESeq.  Note that the 3-way Monte Carlo simulation runs 1000 tests as compared to pairwise DESeq calculations of 3000 p-values.  Adjustment of the 3-way Monte Carlo p-values was done to control the false discovery rate using the Benjamini-Hochberg procedure \cite{BH95}. Following the one-step procedure outlined by Jiang and Doerge \cite{Jia06}, the pairwise p-values were also adjusted to control the false discovery rate as follows.  The three sets of pairwise p-values were combined and the Benjamini-Hochberg procedure was performed on the combined set; subsequently, the adjusted p-values were uncombined and allocated back to the respective sample and gene.  A gene was considered significant if the FDR adjusted p-value (or min(p-value) in the three pairwise comparison setting) was less than 0.05.

\begin{figure}[ht]
\begin{center}
\includegraphics[scale=.5,angle=0]{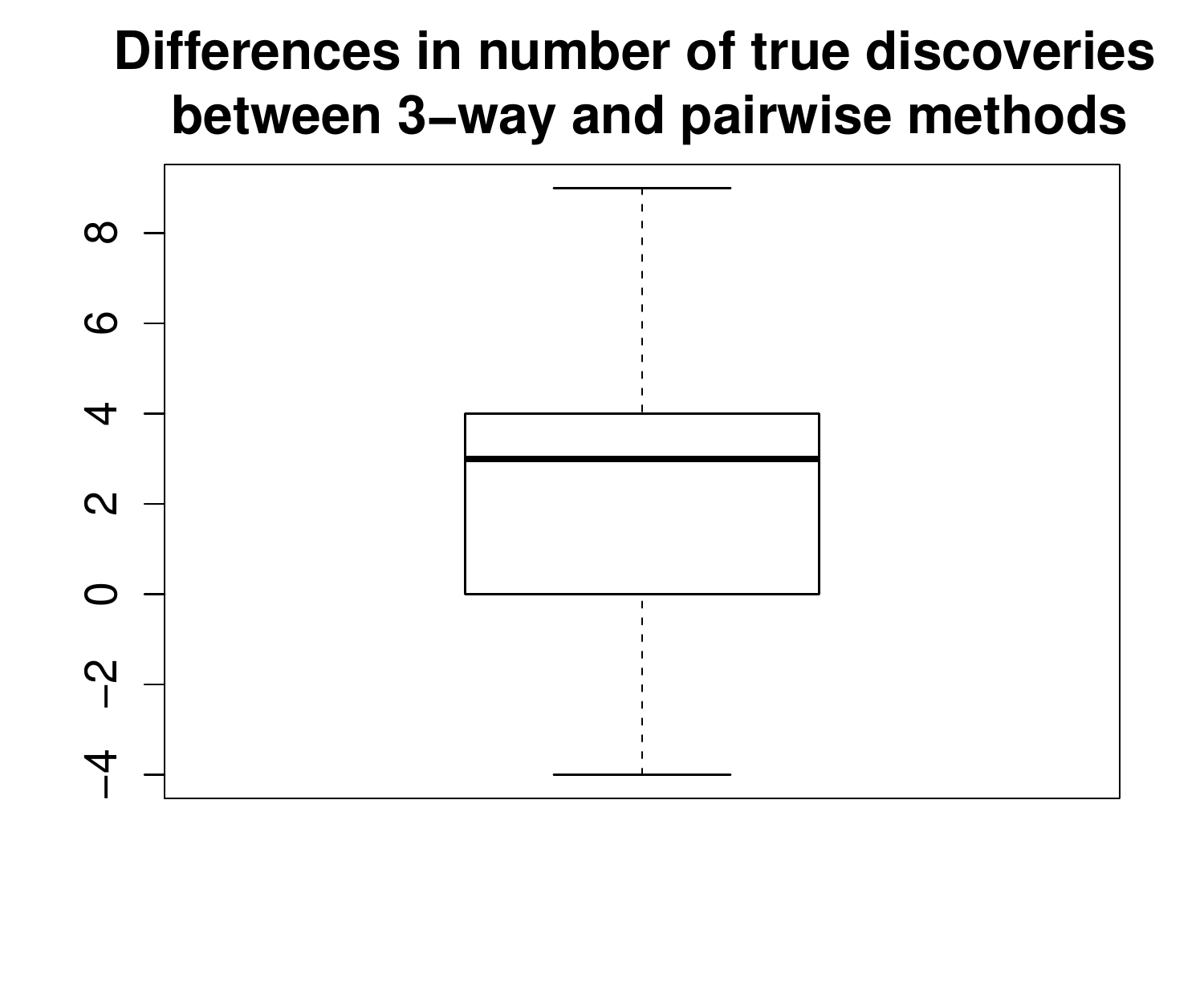}
\end{center}
\caption{\label{truediff.fig} For each of 100 simulations, we measure the increase in number of true discoveries using 3-way Monte Carlo p-values as compared with pairwise DESeq p-values, out of 100 significant genes.  In the majority of the simulations, the 3-way Monte Carlo method was able to find more truly differentially expressed genes than the pairwise comparisons.}
\end{figure}

Table \ref{truediffcount.tab} and Figure \ref{truediff.fig} show the difference in number of significant genes for the 3-way Monte Carlo simulation method as compared to the pairwise DESeq analysis.  Though the increase in power is modest, we emphasize that it is consistent.  We have seen a similar increase in power over many different simulations (data not shown) and with real data (e.g., see section \ref{sectionData}).  As with other global statistics tests (e.g., ANOVA), the modest increase in power seen in this simulation will grow substantially as the number of groups increases.  For example, consider the simulation method above where one group is differentially expressed as compared to the remaining groups.  In the simulation, there are 100 (of 1000) truly differentially expressed genes to find using the 3-way Monte Carlo method; there are 200 (of 3000) truly differentially expressed genes to find using the pairwise DESeq method.  Generally, with $c$ conditions and a similar setup (100 significant genes out of 1000 total genes), the number (and proportion) of truly significant tests will be given as in Table \ref{propdiffcount.tab}.  By considering the number of truly differentially expressed genes as a proportion of the number of tests run, it is clear that as the number of pairwise comparisons grows, the FDR adjustment will lead to non-discovery of more and more truly differentially expressed genes.

\begin{center}
\begin{table}
\begin{tabular}{cccccc}
Min. &  1st Qu.   & Median &    Mean &     3rd Qu. &    Max.\\
 -4  &  0   &   3  &  2.39 &    4 &  9
 \end{tabular}
 \caption{\label{truediffcount.tab}  Summary statistics of the total number of additionally found significant genes out of 100 (in comparing the 3-way Monte Carlo p-values with the DESeq pairwise comparisons.}
 \end{table}
 \end{center}

\begin{center}
\begin{table}
\begin{tabular}{lccc}
method & number DE tests & number tests & prop DE tests\\
\hline
3-way MC & $100$ & 1000 & $\frac{1}{100}$\\
&&&\\
pairwise DESeq & $(c-1) \cdot 100$ & ${c \choose 2} 1000$ & $\frac{(c-1)}{{c \choose 2} 100}$
\end{tabular}
 \caption{\label{propdiffcount.tab}  Table of the number and proportion of truly differentially expressed genes for the two methods as a function of $c$, the number of conditions under study.  Consider the situation with 100 truly differentially expressed genes out of 1000 genes total and one condition differentially expressed as compared to the other $(c-1)$ conditions.}
\end{table}
\end{center}

\subsection{False Discovery Rate}

\begin{figure}
\centering
\begin{subfigure}{.5\textwidth}
  \centering
  \includegraphics[width=.9\linewidth]{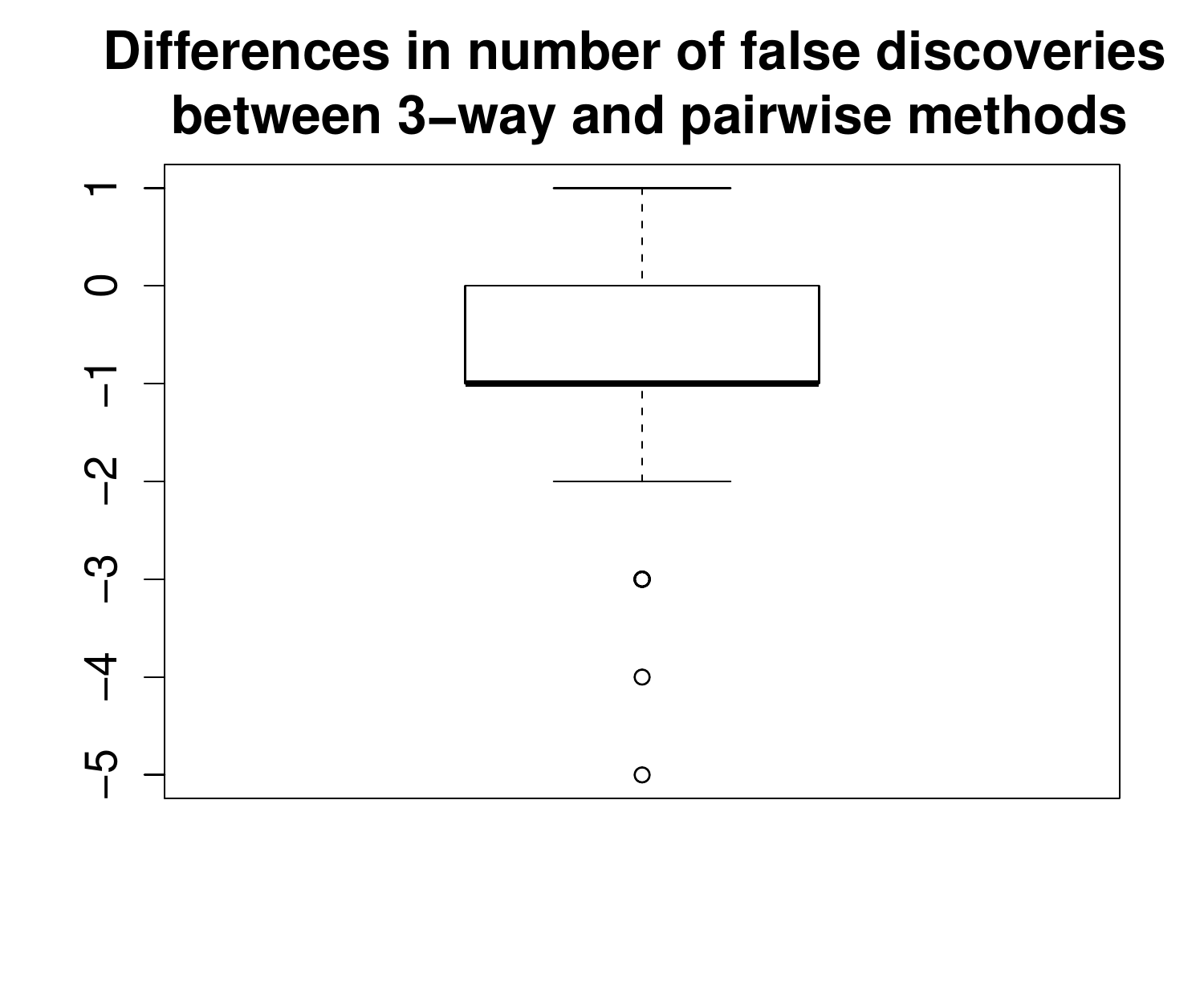}
  \caption{Difference in number of false discoveries}
\end{subfigure}%
\begin{subfigure}{.5\textwidth}
  \centering
  \includegraphics[width=.9\linewidth]{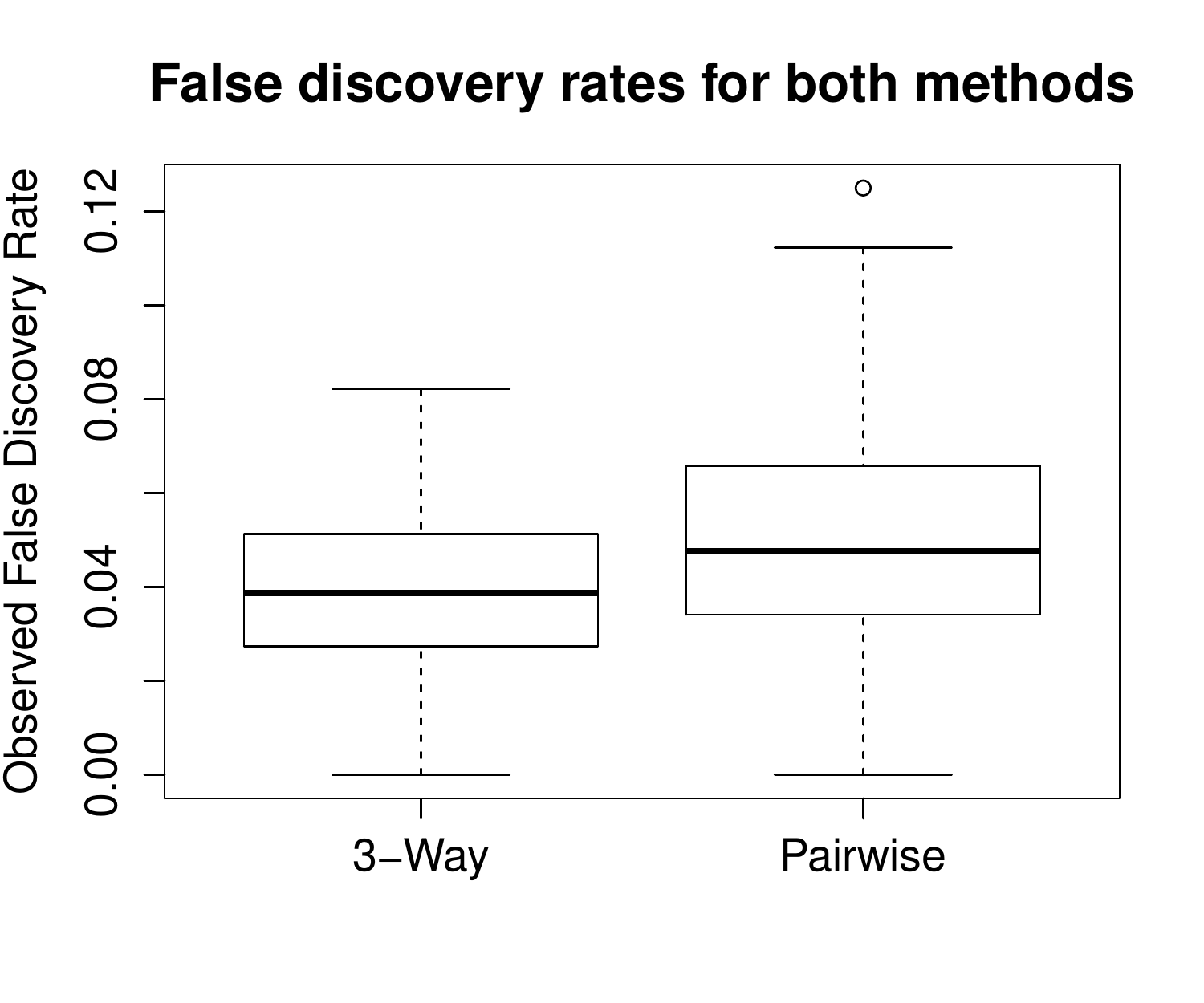}
  \caption{Empirical false discovery rate}
\end{subfigure}
\caption{\label{fdr.fig} For each of 100 simulations, we measure the increase in number of false discoveries using 3-way Monte Carlo p-values as compared with pairwise DESeq p-values, out of 100 significant genes.  In the majority of the simulations, the 3-way Monte Carlo method designated as significant fewer null genes than the pairwise comparisons.}
\end{figure}

Additionally, we compare the empirical false discovery rates for the two methods.   Figure \ref{fdr.fig} shows that for each individual simulation, the pairwise method identifies more null genes as significant more often as compared to the 3-way Monte Carlo procedure.

\section{Experimental Data \label{sectionData}}

We ran the method on publicly available experimental data from Taveirne et al. \cite{Tav13}.  The analysis demonstrates the computing efficiency gained by the Monte Carlo simulation as well as the power improvement over the pairwise comparisons.

\subsection{Biological Experiment}

Our sample data set comes from a study that examined the expression of all genes of the bacterium {\em Campylobacter jejuni} \cite{Tav13}. Campylobacter is a major cause of food poisoning in the US \cite{Alt99}. One major source of infection is the handling of raw poultry, so there is a great deal of interest in understanding how this bacterium lives both in and out of poultry. To do this, Taveirne et al.~\cite{Tav13} studied expression under three conditions. One condition was growth of the bacterium in digestive tract of chickens. The other two were laboratory environments. The first was during active growth (called mid-log), and the second was a non-growing condition called stationary phase. The experiment involved three biological replicates from each condition. Each replicate possessed between 2 to $6 \times 10^7$ reads that mapped back to genes. The experimental samples were used to assess differential gene expression across the three conditions.

\begin{figure}[ht]
\begin{center}
\includegraphics[scale=.6,angle=0]{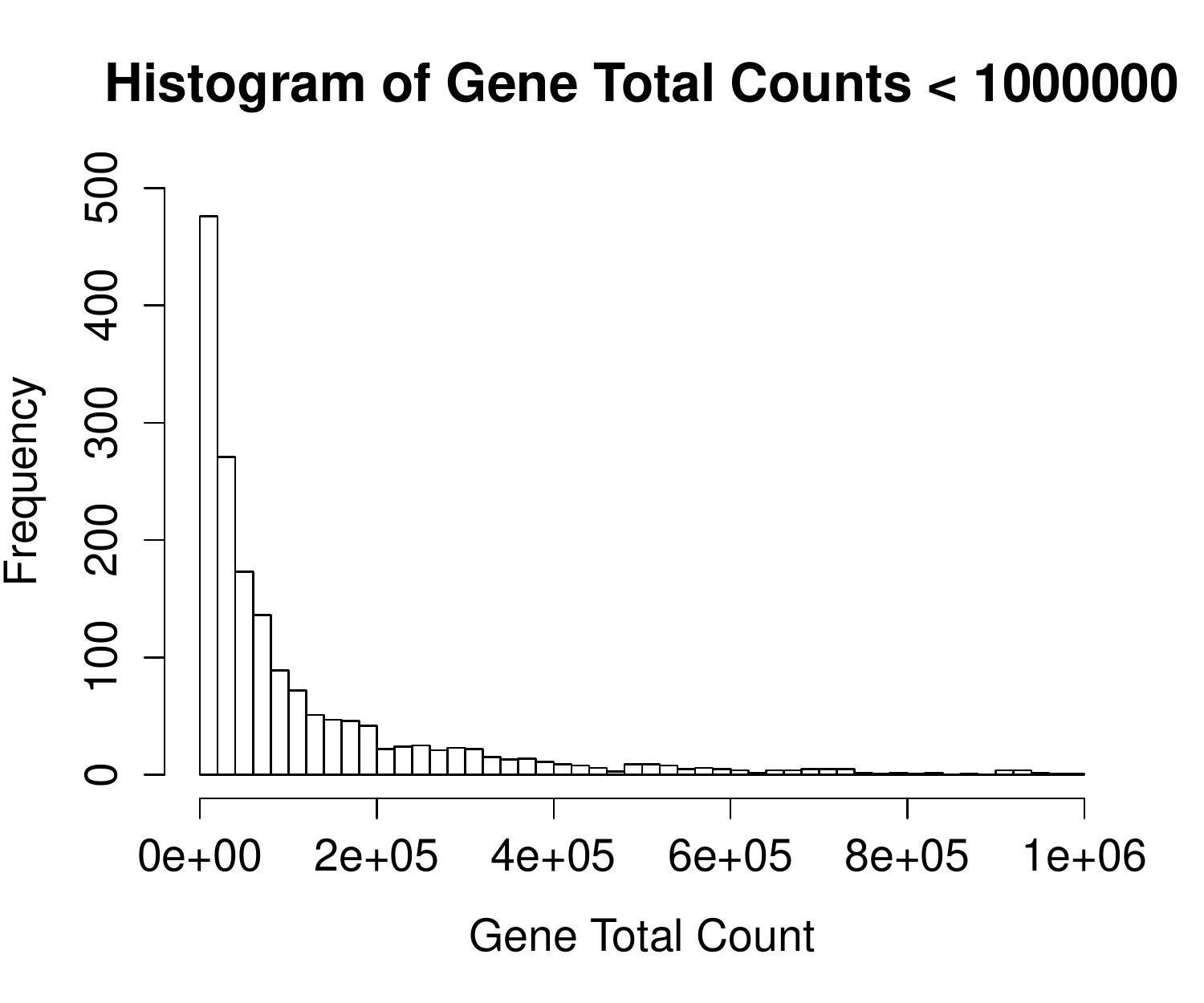}
\end{center}
\caption{\label{totcount.fig} Histogram of the total counts ($k_{is}$) less than one million; produced for the genes in the {\em C. jejuni} dataset.}
\end{figure}

Because the computational limits of the exact extension of DESeq from two to three conditions depends on the total count across all nine sample ($k_{is}$) for each gene, we investigate the total counts from the {\em C. jejuni} data.  Though there are certainly some genes with small total counts, (one quarter of the genes have a total count of 17,920 or fewer), the majority of genes have a count total far above the level where the exact three condition comparison is possible (see Table \ref{totcount.tab} and Figure \ref{totcount.fig}).

\begin{center}
\begin{table}
\begin{tabular}{cccccc}
Min. &  1st Qu.   & Median &    Mean &     3rd Qu. &    Max.\\
 0  &  17920   &   53530  &  193700 &    161700 &  23550000
 \end{tabular}
 \caption{\label{totcount.tab}  Summary statistics of the total counts ($k_{is}$) produced for each of the 1,758 genes in the {\em C. jejuni} dataset.}
 \end{table}
 \end{center}

\subsubsection{Run Time}

\begin{figure}[ht]
\begin{center}
\includegraphics[height=3in, width=6in]{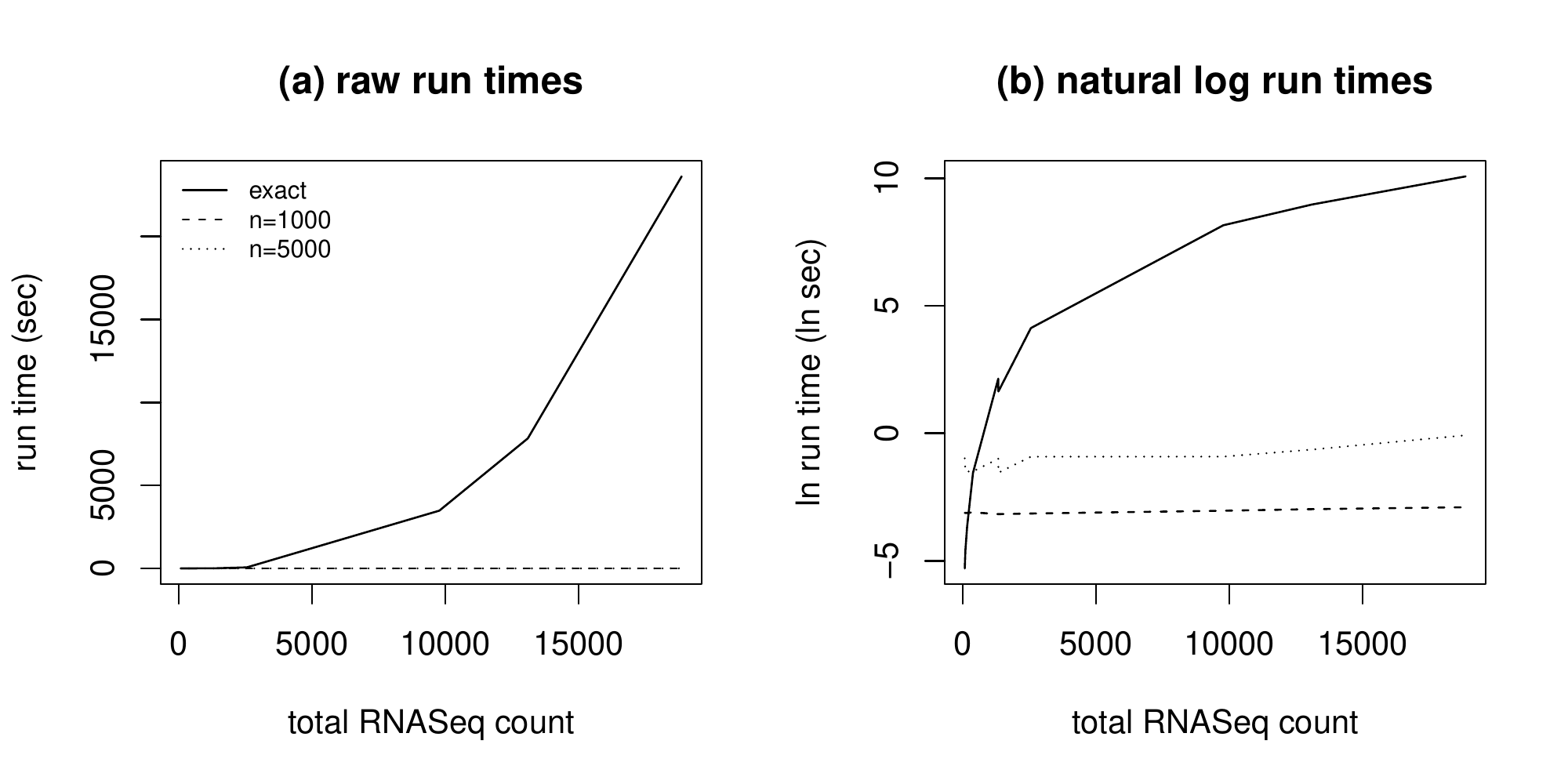}
\end{center}
\caption{\label{runtime} The figure shows the run time as a quadratic function of the total count.  As compared to the Monte Carlo simulation which is linear in total count.  In panel (a) The Monte Carlo simulations for samples of size 1000 and 5000 are indistinguishable as compared to the exact DESeq run time. In panel (b) the run time is natural log transformed, and the run times for the three methods are more easily differentiated.}
\end{figure}

As indicated in Figure \ref{runtime}, the Monte Carlo simulations are substantially faster to run than the exact DESeq p-values across a three group comparison.  Using a few of the (smaller) total gene count values from the {\em C. jejuni} data \cite{Tav13}, we calculated the run times associated with exact DESeq p-values, Monte Carlo p-values from samples of size 1000, and Monte Carlo p-values from samples of size 5000.  The analysis was performed on a computer with two eight core AMD Opteron 6276 processors running at 1.4 GHz.

\subsubsection{Power}

To examine the performance of the 3-way Monte Carlo DESeq method we analyzed the full Taveirne et al. {\em C. jejuni} dataset. The 3-way Monte Carlo DESeq method was used to calculate p-values for differential expression across all three conditions simultaneously, while the standard DESeq package was used to calculate p-values for each of three pairwise group comparisons. As with the simulated data, adjustment of the 3-way Monte Carlo DESeq p-values was done using the Benjamini-Hochberg procedure \cite{BH95}. Following the one-step procedure outlined by Jiang and Doerge \cite{Jia06}, the three sets of pairwise p-values were combined and the Benjamini-Hochberg procedure was performed on the combined set. As in Taveirne et al., a gene was determined to be significant under the 3-way Monte Carlo DESeq procedure if the largest fold change between a pair of conditions was at least 4 (ratio of average expression across two groups was greater than 4), and the FDR-adjusted p-value for differential expression was below 0.05. Similarly, under the standard DESeq procedure we called a gene significant if the largest fold change was at least 4 and the minimum adjusted pairwise p-value was below 0.05. The 3-way Monte Carlo DESeq method declared 344 genes significant while the pairwise DESeq method found 342. While the difference in the two methods seems minimal, it is important to note that {\em all genes declared significant by the pairwise method were also declared significant under the Monte Carlo method.}  Additionally, the number of significant genes found is consistent with Taveirne et al.

The analysis was performed on a computer with two eight core AMD Opteron 6276 processors running at 1.4 GHz in just under one hour.  Note that calculating the $p$-value exactly for just the gene with $23 \cdot 10^6$ counts would require on the order of $10^{12}$ floating point operations, making this infeasible in practice.

\section{Conclusions}

We extend DESeq to a three or more simultaneous group comparison. The time to compute an exact $p$-value for $m$ biological conditions is $\Theta(k_{iS}^{m - 1})$. Since actual data sets can have total counts for genes ($k_{iS}$) of $10^6$ or more, the exact calculation is infeasible even for three biological conditions.  Our method requires a thoughtful sampling scheme but is otherwise straightforward to apply using the built in DESeq functions.  Additionally, we are able to show that the pairwise exact DESeq comparisons give a subset significant genes as compared to the three group comparison using 3-way Monte Carlo simulations.  As seen with the simulated data and the data from Taveirne et al., the three group comparison appears to be more powerful than the pairwise comparisons and just as computationally accessible.

\section{Acknowledgements}
Work was supported by Howard Hughes Medical Institute Undergraduate Science Education Program awards \#52006301 to Harvey Mudd College and
\#5200755 to Pomona College.

\bibliography{differential}{}
\bibliographystyle{abbrvnat}

\end{document}